\documentstyle[prl,aps,twoside,multicol]{revtex}

\def\CC{{\rm\kern.24em \vrule width.04em height1.46ex depth-.07ex 
\kern-.30em C}}

\begin{document}
\title{Universal Fault-Tolerant Computation on Decoherence-Free Subspaces}
\author{D. Bacon$^{1,2}$, J. Kempe$^{1,3,4}$, D.A. Lidar$^{1}$ and
K.B. Whaley$^{1}$} 
\address{Departments of Chemistry$^{1}$, Physics$^{2}$ and
Mathematics$^{3}$,\\
University of California, Berkeley\\
\'{E}cole Nationale Superieure des T\'{e}l\'{e}communications,\\
Paris, France$^4$}
\date{\today}
\maketitle

\begin{abstract}
A general scheme to perform universal quantum computation 
within decoherence-free subspaces (DFSs) of a system's Hilbert space is
presented. This scheme leads to the first fault-tolerant realization of
universal quantum computation on DFSs with the properties that (i) only one-
and two-qubit interactions are required, and (ii) the system remains within
the DFS throughout the entire implementation of a quantum gate. We show
explicitly how to perform universal computation on clusters of the
four-qubit DFS encoding one logical qubit each under ``collective
decoherence'' (qubit-permutation-invariant system-bath coupling). Our
results have immediate relevance to a number of solid-state quantum computer
implementations, in particular those in which quantum logic is
implemented through exchange interactions, such as the recently proposed
spin-spin coupled GaAs quantum dot arrays and the Si:$^{31}$P nuclear
spin arrays.
\end{abstract}


\begin{multicols}{2}

Decoherence-free subspaces (DFSs) have been recently proposed \cite
{Zanardi:97,Duan:98,Lidar:PRL98} to protect fragile quantum
information against the detrimental effects of decoherence. This is
especially important for quantum computation, where maintaining the quantum
coherence of states forms the cornerstone of the promised speed-up compared
to classical computers \cite{Preskill:97b}. Under certain assumptions
about the symmetry of the noise processes, most notably spatially correlated
errors, there exist subspaces of the system's Hilbert-space that are not
affected by the noise, and are thus decoherence-free.
Maintaining a system inside a DFS can therefore be thought of as a
``passive'' error-{\em prevention} scheme. It was shown recently that DFSs
are robust under symmetry-breaking perturbations and are thus ideal codes
for quantum {\em memory} \cite{Bacon:99}. So far it was not known whether
quantum {\em computation} on DFSs is possible without catastrophically
taking the system outside the DFS [thus exposing it to (collective) errors]
under realistic physical constraints. Realistic implementable gates are
restricted to one- and two-body interactions and a finite number of
measurements. Previous demonstrations of universal quantum computation on
DFSs did not satisfy these criteria \cite{Lidar:PRL98,Zanardi:99a}.

In this work we develop a formalism that allows us to find Hamiltonians
involving only one- and two-qubit interactions, which can be used to
implement universal quantum gates {\em without ever leaving the
DFS}. When computation is performed in this manner the system is never
exposed to 
errors, so that this approach is {\em naturally fault-tolerant}. This
is in marked contrast to ``active'' quantum error correction codes
(QECCs) \cite
{Calderbank:96Steane:96a,Knill:97b}, where errors do take codewords out of the codespace, and
fault-tolerance requires a hefty overhead \cite{Aharonov:99a}. Moreover, when the
symmetric noise process allowing for the existence of the DFS is perturbed,
we show that it is possible nevertheless to completely stabilize the
computation, fault-tolerantly, by using a concatenation scheme with a QECC. This scheme, which we first proposed in 
\cite{Lidar:PRL99}, has the advantage that it operates with an
error-threshold that depends only on the {\em perturbing} error rate, and is
thus particularly attractive considering the stringent requirement on the
threshold for fault-tolerant quantum computation \cite{Aharonov:99a}. The
formalism we develop here connects DFSs with the theory of stabilizer QECCs 
\cite{Gottesman:97}. We apply it to derive gates in the important {\em 
collective decoherence} model (the decoherence mechanism expected to dominate
in the solid state at very low temperatures), and explicitly construct a
universal set of gates operating on clusters of $4$ physical qubits that
each encode one logical qubit. In conjuction with the concatenated code of 
\cite{Lidar:PRL99}, this suffices to implement universal fault-tolerant
computation on DFSs {\em robustly}. Thus we show here that one can employ the full power of
DFSs in preserving coherence not merely for quantum memory
applications, but also for full-scale quantum computing. We outline an
application of our results to a class of potential physical implementations
of quantum computers, in which quantum logic is implemented through
internal exchange interactions. This class includes the recently
proposed spin-spin coupled 
quantum dot arrays \cite{LossBurkard:98} and the silicon-based nuclear
quantum computer \cite{Kane:98}.

{\it Conditions for decoherence-free subspaces.}--- Consider the dynamics of
a system $S$ coupled to a bath $B$ which evolves unitarily under the
combined system-bath Hamiltonian ${\bf H}={\bf H}_{S}\otimes {\bf I}_{B}+
{\bf I}_{S}\otimes {\bf H}_{B}+\sum_{\alpha =1}^{A}{\bf S}_{\alpha }\otimes 
{\bf B}_{\alpha }$, where ${\bf H}_{S}$ (${\bf H}_{B}$) is the system (bath)
Hamiltonian, ${\bf I}_{S}$ (${\bf I}_{B}$) is the identity operator on the
system (bath), and ${\bf S}_{\alpha }$ (${\bf B}_{\alpha }$) acts solely on
the system (bath). The last term in ${\bf H}$ is the interaction Hamiltonian 
${\bf H}_{I}$. The evolution in a subspace $\tilde{{\cal H}}$ of the system
Hilbert space ${\cal H}$ is {\em unitary} for all possible bath states iff
(i) 
\begin{equation}
{\bf S}_{\alpha }|\psi \rangle =c_{\alpha }|\psi \rangle \quad c_{\alpha
}\in \CC
\label{eq:DFS}
\end{equation}
for all states $|\psi \rangle $ which span $\tilde{{\cal H}}$, and for every
operator ${\bf S}_{\alpha }$ in ${\bf H}_{I}$, (ii) $S$ and $B$ are
initially decoupled, and (iii) ${\bf H}_{S}|\psi \rangle $ has no overlap
with states in the subspace orthogonal to $\tilde{{\cal H}}$ \cite
{Zanardi:97,Lidar:PRL99}. A subspace of ${\cal H}$ which fullfills these
requirements is a {\em decoherence-free subspace} (DFS). It is important to
notice that if condition (iii) is not fullfilled then states leak out of the
DFS, and the usefulness of these subspaces for the storage of quantum
information is lost.

{\it General stabilizer formalism.}--- In order to identify a set of
fault-tolerant universal gates for computation on a DFS, and to make a more
explicit connection to QECCs, we recast the definition of a DFS [Eq.~(\ref
{eq:DFS})] into the {\em stabilizer formalism}. By analogy to QECC \cite
{Gottesman:97}, we define the {\em DFS stabilizer} ${\cal S}$ as a set of
operators ${\bf D}_{\beta }$ which act as identity on the DFS states: 
\begin{equation}
{\bf D}_{\beta }|\psi \rangle =|\psi \rangle \quad \forall {\bf D}_{\beta
}\in {\cal S}\quad {\rm iff}\,\quad |\psi \rangle \in {\rm DFS}.
\label{eq:superdef}
\end{equation}
Here $\beta $ can be a discrete or continuous index; ${\cal S}$ can form a
finite set or group. Our ${\cal S}$ is a generalization of the QECC
stabilizers which restrict ${\cal S}$ to an Abelian subgroup of the Pauli
group (the group formed by tensor products of Pauli matrices on the qubits) 
\cite{Gottesman:97}. While some DFSs can also be specified by a stabilizer
in the Pauli-group \cite{Lidar:PRA99Pauli}, many DFSs are specified by
non-Abelian groups, and hence are {\em nonadditive} codes \cite{Rains:97}.
The stabilizer of a QECC allows identification of the errors the code can
correct. In general an error-process can be described by the Kraus
operator-sum formalism \cite{Knill:97b}: $\rho \rightarrow \sum_{\mu }{\bf A}
_{\mu }\rho {\bf A}_{\mu }^{\dagger }$. The Kraus-operators ${\bf A}_{\mu }$
can be expanded in a basis ${\bf E}_{i}$ of ``errors''. Two types of errors
can be dealt with by stabilizer codes: (i) errors ${\bf E}_{i}^{\dagger }
{\bf E}_{j}$ that anticommute with any ${\bf S}\in {\cal S}$, and (ii)
errors that are part of the stabilizer (${\bf E}_{i}\in {\cal S}$). The
first class are errors that require active correction; the second class (ii)
are ``degenerate'' errors that do not affect the code at all. A duality
between QECCs and DFSs can be stated as follows: QECCs were designed
primarily to deal with type (i) errors, but can also be regarded as DFSs for
the errors in their stabilizer \cite{Lidar:PRA99Pauli}. Conversely, DFSs
were designed primarily to deal with type (ii) errors, but can in
principle be used as a QECC against errors that are type (i) with
respect to ${\cal S}$.

Consider now the following continuous index stabilizer: 
\begin{equation}
{\bf D}(v_{0},v_{1},...,v_{A})={\bf D}(\vec{v})=\exp \left[ \sum_{\alpha
=1}^{A}\left( c_{\alpha }{\bf I}-{\bf S}_{\alpha }\right) v_{\alpha }\right]
.  \label{eq:DFSstab}
\end{equation}
Clearly, the DFS condition [Eq.~(\ref{eq:DFS})] implies that ${\bf D}(\vec{v})|\psi \rangle =|\psi \rangle $. Conversely, if ${\bf D}(\vec{v})|\psi
\rangle =|\psi \rangle $ for all $\vec{v}$, then in particular it must hold
that for each $\alpha $, $\exp \left[ \left( c_{\alpha }{\bf I}-{\bf S}
_{\alpha }\right) v_{\alpha }\right] |\psi \rangle =|\psi \rangle $.
Recalling that $\phi ({\bf A)=}\exp {\bf [A}]$ is a one-to-one continuous
mapping of a small neighbourhood of the zero matrix ${\bf 0}$ onto a small
neighbourhood of the identity matrix ${\bf I}$, it follows that there must
be a sufficiently small $v_{\alpha }$ such that $\left( c_{\alpha }{\bf I}-
{\bf S}_{\alpha }\right) |\psi \rangle =0$. Therefore the DFS condition (\ref
{eq:DFS}) holds iff ${\bf D}(\vec{v})|\psi \rangle =|\psi \rangle $ for all $
\vec{v}$.

In order to achieve general-purpose {\em universal} quantum computation one
must demonstrate that one can perform a set of operations (gates) ${\bf U}$
which allow for the implementation of nearly every unitary operation on the
quantum computer (dense in the set of unitary operations) \cite{Lloyd:95}.
In analogy to computation using physical qubits, for universal computation
on DFSs two types of gates will be needed: (i) gates performing operations
within a DFS; and (ii) gates linking two or more DFS-{\em clusters} (thus
performing operations between logical qubits encoded into different
clusters). The stabilizer formalism is useful for identifying allowed gates
that take codewords to codewords \cite{Gottesman:97}. The {\em allowed}
operations ${\bf U}$ transform the stabilizer into itself. Let $|\psi
\rangle \in \tilde{{\cal H}}$, i.e., ${\bf D}(\vec{v})|\psi \rangle =|\psi
\rangle $. For ${\bf U}$ to be an allowed operation, ${\bf U}|\psi \rangle $
must be in $\tilde{{\cal H}}$, so ${\bf D}(\vec{v}^{\prime })\left( {\bf U}
|\psi \rangle \right) ={\bf U}|\psi \rangle $. This means ${\bf U}{\bf D}(
\vec{v}){\bf U}^{\dagger }={\bf D}(\vec{v}^{\prime }(\vec{v}))$ and the $
{\bf D}(\vec{v}^{\prime }(\vec{v}))$ must cover ${\cal S}$. It is sufficient
to have $\vec{v}^{\prime }(\vec{v})$ to be a one-to-one mapping. If ${\cal S}
$ is a unitary group then the set of allowed gates is the {\em normalizer}
of ${\cal S}$ \cite{Gottesman:97a}. To derive a similar condition for (ii)
which involves gates between two different DFS-clusters with stabilizers $
{\cal S}_{1}$ and ${\cal S}_{2}$, we note that ${\cal S}_{12}={\cal S}
_{1}\otimes {\cal S}_{2}$ is a stabilizer for the two DFS-clusters. The
gates ${\bf U}$ are unitary transformations performed by switching on
Hamiltonians ${\bf H}$ for some time $t$, acting on physical qubits in the
DFS. So far we only required that the action of the gate preserve the
subspace at the conclusion of the gate operation, but not that the subspace
be preserved throughout the entire duration of the gate operation. By posing
the stronger requirement that {\em the state of the system stays inside the
DFS during the entire switching-time of the gate} we achieve {\em natural
fault-tolerance} on the DFS. Rewriting our condition as ${\bf U}(t){\bf D}(
\vec{v})={\bf D}(\vec{v}^{\prime }(\vec{v})){\bf U}(t)$, taking the
derivative with respect to $t$ and evaluating this at $t=0$ we obtain:

{\it Theorem:} A sufficient condition for the generating Hamiltonian to keep
the state at all times entirely within the DFS is ${\bf H}{\bf D}(\vec{v})=
{\bf D}(\vec{v}^{\prime }(\vec{v})){\bf H}$ where $\vec{v}^{\prime }(\vec{v})
$ is one-to-one and time-independent.

{\it Collective Decoherence.}--- We now focus on a particularly important
system-bath interaction model, in which clusters of several qubits couple to
the same bath mode: the {\em collective decoherence} model. Specifically,
the interaction Hamiltonian is of the form ${\bf H}_{I}=\sum_{\alpha =x,y,z}
{\bf S}_{\alpha }\otimes {\bf B}_{\alpha }$, where ${\bf S}_{\alpha }\equiv
\sum_{j=1}^{K}{\bf \sigma }_{\alpha }^{j}$ and $\sigma _{\alpha }^{j}$ are
the Pauli-matrices applied to the $j^{{\rm th}}$ qubit. The ${\bf S}_{\alpha
}$ form the (semi-simple) Lie algebra $su(2)$ and the DFS condition (\ref
{eq:DFS}) becomes ${\bf S}_{\alpha }|\psi \rangle =0$ \cite{Lidar:PRL98}.
The stabilizer for the collective decoherence DFS [Eq.~(\ref{eq:DFSstab})]
is given by 
\[
{\bf D}(\vec{v})=e^{i\vec{v}\cdot {\bf \vec{S}}}=\bigotimes_{j=1}^{K}\left[ 
{\bf I}^{j}\cos ||\vec{v}||+\vec{\sigma}^{j}\cdot \vec{v}/||\vec{v}||\sin ||
\vec{v}||\right] 
\]
where $\vec{{\bf S}}=({\bf S}_{x},{\bf S}_{y},{\bf S}_{z})$ and $||\vec{v}
||\equiv (\sum_{\alpha }v_{\alpha }^{2})^{1/2}$ may be complex. ${\bf D}(
\vec{v})$ consists of all collective qubit rotations+contractions, i.e. an
operation of the form ${\bf G}^{\otimes K}$ where ${\bf G}(\vec{v})$ is any
element in $SL(2)$. The smallest number of physical qubits yielding a full
``encoded DFS qubit'' is four \cite{Zanardi:97}, given by the two
states with $0$ total angular momentum: 
\begin{eqnarray}
|0_{L}\rangle  &=&|s\rangle \otimes |s\rangle   \nonumber \\
|1_{L}\rangle  &=&\frac{1}{\sqrt{3}}[|t_{+}\rangle \otimes |t_{-}\rangle
-|t_{0}\rangle \otimes |t_{0}\rangle +|t_{-}\rangle \otimes |t_{+}\rangle ],
\label{eq:4qubit}
\end{eqnarray}
where $|s\rangle =(|01\rangle -|10\rangle )/\sqrt{2}$ is the singlet state
of two qubits, and $|t_{-,0,+}\rangle =\{|00\rangle ,(|01\rangle +|10\rangle
)/\sqrt{2},|11\rangle \}$ are the corresponding triplet states. Let $\tilde{
{\cal H}}_{4}=\{|0_{L}\rangle ,|1_{L}\rangle \}$; $\tilde{{\cal H}}_{4}$ is
immune to all errors which can be written as sums of collective operations
of the form ${\bf G}^{\otimes 4}$. In addition it is easy to check that $
\tilde{{\cal H}}_{4}$ is a distance-2 QECC \cite{Knill:97b,Kempe:00},
meaning that it can {\em detect} arbitrary single-qubit errors.

{\it Universal gates for a logical qubit.}--- We now apply the general
formalism developed above to derive a set of universal gates on clusters of $
4$-qubit DFSs under collective decoherence. While it is certainly desirable
to consider computation using DFSs over arbitrary block size \cite{Kempe:00}
, it is important to realize that the encoding into $L=K/4$ blocks of four
is entirely sufficient to implement universal ``encoded quantum
computation'' over $\tilde{{\cal H}}_{4}^{\otimes L}$. The $4$-qubit DFS is
also of special interest in the concatenation-scheme proposed in \cite
{Lidar:PRL99}. There the DFS qubit of Eq.~(\ref{eq:4qubit}) becomes the
building block for a QECC that protects against perturbing single physical
qubit errors. By using DFS qubits instead of physical qubits, the resulting
concatenated code offers an error threshold for faul-tolerant computation
that depends only on the {\em perturbing} single qubit error rate. However,
the threshold for QECC relies heavily on the ability to perform error
correction (i.e. ``computation'') which does not catastrophically create
more errors than it fixes (fault-tolerance). Concatenated DFS-QEC-codes will
only be efficient if (i) a realistic (no more than $2$-body interactions)
set of universal quantum gates keeping states entirely {\em within} the DFS
is used, and (ii) preparation and decoding of DFS states is performed
fault-tolerantly. Here we provide such a set of universal quantum gates, and
detail the preparation and decoding of DFS states.

It is sufficient to be able to apply (i) all single qubit rotations [$SU(2)$] together with (ii) the two-qubit controlled phase gate (${\bf C}_{P}$
defined below) on any two logical qubits, in order to perform any unitary
transformation \cite{DiVincenzo:95}. We now show how to construct this
universal set of gates. The stabilizer for the $4$-qubit DFS $\tilde{{\cal H}
}_{4}$ is of the form ${\bf D}(\vec{v})={\bf G}^{\otimes 4}$, which is
manifestly invariant under permutations of the qubits. The Hermitian
exchange (transposition) operation that switches only qubits $i$ and $j$, $
{\bf E}_{ij}|x\rangle _{i}|y\rangle _{j}=|y\rangle _{i}|x\rangle _{j}$ ($
x,y=0\,$\ or $1$), leaves the stabilizer element-wise invariant and so
trivially fulfills the conditions of the Theorem (with $\vec{v}^{\prime }=
\vec{v}$). Thus $\exp [-i\theta {\bf E}_{ij}]$ preserves the DFS, and is a
valid unitary operation by a two physical qubit Hamiltonian [see also \cite
{Lidar:PRA99Exchange}]. Consider the action of ${\bf E}_{ij}$ on the basis
states of Eq.~(\ref{eq:4qubit}): ${\bf E}_{12}={\bf E}_{34}$ acting on these
states takes $|0_{L}\rangle \rightarrow -|0_{L}\rangle $ and $|1_{L}\rangle
\rightarrow |1_{L}\rangle $. $-{\bf E}_{12}$ thus acts as the encoded $
\sigma _{z}$ ($\bar{Z}$ -- a bar indicates operations on the
encoded DFS-qubits). Furthermore, ${\bf E}_{13}={\bf E}
_{24}$ applied to the basis states in Eq.~(\ref{eq:4qubit}) acts as 
\begin{equation}
{\bf E}_{13}=\left( 
\begin{array}{cc}
{\frac{1}{2}} & -{\frac{\sqrt{3}}{2}} \\ 
-{\frac{\sqrt{3}}{2}} & -{\frac{1}{2}}
\end{array}
\right) .
\end{equation}
Thus ${\bf H}_{x}=-{\frac{2}{\sqrt{3}}}({\bf E}_{13}+{\frac{1}{2}}{\bf E}
_{12})$ acts as an encoded $\sigma _{x}$ ($\bar{X}$) on the DFS qubit. ${\bf 
H}_{x}$ can be implemented either by turning on the two 2-qubit Hamiltonians
at the same time, or approximated by using a finite number of terms in
the Lie sum 
formula: $e^{i(\alpha A+\beta B)}=\lim_{n\rightarrow \infty }\left(
e^{i\alpha A/n}e^{i\beta B/n}\right) ^{n}=e^{i\alpha A/n}e^{i\beta
B/n}+O(1/n^{2})$. The ability to implement $\bar{Z}$ and $\bar{X}$ is
sufficient to complete the Lie algebra $\overline{su(2)}$, and thus to
implement any gate in $SU(2)$ on the encoded qubits. This can be done using
an Euler angle construction for the desired gate as is done routinely
in NMR \cite{Slichter:96}, or, following the standard
arguments for universality \cite{DiVincenzo:95}, through an
approximation to the Lie product
formula $e^{[A,B]}=\lim_{n\rightarrow \infty } Q^n = Q +
O(1/n^{3/2})$, where $Q = e^{iA/\sqrt{n}}e^{iB/
\sqrt{n}}e^{-iA/\sqrt{n}}e^{-iB/\sqrt{n}}$. We note that
the Euler angle construction is in general simpler to implement in
practical implementations because it does not require extremely fast
switching, but defer the construction of optimal gate sequences to 
future work. To conclude, {\em we can
generate any $SU(2)$ operation on the encoded qubits by simply turning on
and off the appropriate 2-qubit exchange Hamiltonian.} 

To complete the universal set of gates we explicitly construct an encoded
controlled-phase gate $\bar{{\bf C}}_{P}$ between two DFS-qubits (i.e. two
separate $4$-qubit DFS-clusters). In doing so we assume that the qubits are
physically close during the gate switching time, so that they form an
$8$-qubit, $14$-dimensional collective-decoherence DFS $\tilde{{\cal
H}}_{8}$ 
(see \cite{Lidar:PRA99Exchange} for a derivation of a general dimension
formula). Four of these $14$ dimensions are spanned by the two $4$-qubit
DFSs ($\tilde{{\cal H}}_{4}\otimes \tilde{{\cal H}}_{4}$). Given two
clusters of $K$ physical qubits each, the exchange
interaction between any two qubits preserves $\tilde{{\cal H}}_{2K}$,
since the stabilizer is just ${\bf G}^{\otimes 2K}$. Thus a
sequence of gates constructed purely out of exchange operations will {\em 
never} take the system out of $\tilde{{\cal H}}_{2K}$. Remarkably, exchange
interactions alone can be shown to generate the full (special) unitary
group on any collective DFS $\tilde{\cal H}_{2K}$ (in particular
$K=4$) \cite{Kempe:00}. Therefore generating $\bar{{\bf C} 
}_{P}$ is a matter of finding an appropriate explicit construction, which we
now present. Defining ${\bf h}_{1}=[{\bf E}_{26},{\bf E}_{12}+{\bf E}_{25}]+[
{\bf E}_{15},{\bf E}_{12}+{\bf E}_{16}]$, ${\bf h}_{2}=\sum_{j=5}^{8}\left( 
{\bf E}_{1j}+{\bf E}_{2j}\right) $, and ${\bf c}={\frac{1}{32}}[{\bf h}_{1},[
{\bf h}_{2},{\bf h}_{1}]]$, a calculation shows that ${\bf c}$ acts on the
two $4$-qubit DFSs as: $|0_{L}0_{L}\rangle \rightarrow 0$, $
|0_{L}1_{L}\rangle \rightarrow |0_{L}1_{L}\rangle $, $|1_{L}0_{L}\rangle
\rightarrow 0$, and $|1_{L}1_{L}\rangle \rightarrow 0$. The Hamiltonian $
{\bf c}$ then yields a controlled phase gate by exponentiation: $\bar{{\bf C}
}_{P}(\theta )=\exp [i{\bf c}\theta ]$. The action of ${\bf c}$ can be
understood as follows: (i) ${\bf h}_{1}$ takes states from $\tilde{{\cal H}}
_{4}\otimes \tilde{{\cal H}}_{4}$ into $\tilde{{\cal H}}_{8}$; (ii) ${\bf h}
_{2}$ then applies a phase to a single one of the states in $\tilde{{\cal H}}
_{8}$; (iii) ${\bf h}_{1}$ returns the states from $\tilde{{\cal H}}_{8}$
into $\tilde{{\cal H}}_{4}\otimes \tilde{{\cal H}}_{4}$. {\em We have thus
explicitly constructed a naturally fault-tolerant universal set of gates,
which utilizes only two-body exchange interactions}.

It remains to be shown that it is possible to prepare and decode states in
the DFS fault tolerantly. We first note that the $|0_{L}\rangle $ state of
Eq.~(\ref{eq:4qubit}) can be constructed by simply preparing two pairs of
qubits in the singlet state (in general for $K\geq 4$ qubits the state $
|0_{L}\rangle $ can always be chosen to be a product of singlets). All the
other DFS-states (like $|1_{L}\rangle $) can be obtained by applying the
appropriate encoded operation to $|0_{L}\rangle $ (like $\bar{X}$). To
verify correct state-preparation and to decode we need {\em fault-tolerant}
measurements in the encoded ``computational'' basis \{$|0_{L}\rangle
,|1_{L}\rangle $\} (eigenstates of $\bar{Z}$). It is easily checked that
measuring $\{\sigma _{z}^{1}$, $\sigma _{z}^{2}$, $\sigma _{x}^{3}$, $\sigma
_{x}^{4}\}$ on the four qubits allows to distinguish $|0_{L}\rangle $ and $
|1_{L}\rangle $ but destroys the DFS-state. To perform a fault-tolerant and
non-destructive measurement of $\bar{Z}$ we require ancilla states prepared
in the DFS-state $|0_{L}\rangle $ and an encoded controlled-not ($\bar{{\bf C
}}_{X}$) gate, which we have at our disposal from the construction of
universal gates above ($\bar{{\bf C}}_{X}$ acts as $|x_{L},y_{L}\rangle
\rightarrow |x_{L},(x_{L}+y_{L}){\rm mod}2\rangle $, where $x,y=0$ or $1$).
By applying a $\bar{{\bf C}}_{X}$ gate between the DFS state to be measured
and the ancilla, and performing a destructive measurement on the ancilla, we
obtain a nondestructive measurement of $\bar{Z}$, which is tolerant of
collective errors on the two DFSs. To prevent possible uncontrolled
error-propagation caused by an incorrectly prepared ancilla, we prepare
multiple $|0_{L}\rangle $-ancillas and apply $\bar{{\bf C}}_{X}$'s between
the DFS state to be measured and each ancilla. Together with majority voting
this provides a fault-tolerant method for measuring $\bar{Z}$ \cite
{Gottesman:97a}. This procedure can also be used to verify the preparation
of $|0_{L}\rangle $, and thus assures fault-tolerant DFS-state preparation.

{\it Application to solid-state quantum computer implementations}.--- Two of
the most promising proposals for quantum computer implementations, the
spin-spin coupled quantum dots \cite{LossBurkard:98} and the Si:$^{31}$P
nuclear spin array \cite{Kane:98}, rely on controllable exchange
interactions for the implementation of quantum logic. The pertinent part of
the internal Hamiltonian is of the Heisenberg type: ${\bf H}_{{\rm Heis}}=
\frac{1}{2}\sum_{i\neq j}J_{ij}{\bf S}_{i}\cdot {\bf S}_{j}.$ Here ${\bf S}
_{i}=(\sigma _{x}^{i},\sigma _{y}^{i},\sigma _{z}^{i})$ is the Pauli matrix
vector of spin $i$ and $J_{ij}$ are exchange-coefficients,
tunable by variation of external parameters such as local electric and
magnetic fields. It is easily checked that ${\bf E}_{ij}\equiv \frac{1}{2}
\left( {\bf I}_{S}+{\bf S}_{i}\cdot {\bf S}_{j}\right) $ is an exchange
operator of physical qubits $i$ and $j$
\cite{Ruskai:99}. Details of the tuning of the $J_{ij}$ were worked out in \cite{LossBurkard:98,Kane:98}, and show high
sensitivity to externally applied electric and magnetic fields. A range of
about 0-1meV is attainable in quantum dots \cite{LossBurkard:98} by
tuning the magnetic field through $0-2$tesla. {\em 
Thus ${\bf H}_{{\rm Heis}}$ is a sum over exchange terms with tunable
coefficients, and can be used to implement quantum computation over a DFS as
detailed above}. A magnetic field $\ge 2$tesla and a temperature $\le 100$mK are required in the
Si:$^{31}$P proposal in order that the electrons occupy only the
lowest energy bound 
state at the $^{31}$P donor \cite{Kane:98}. At these extremely low
temperatures we expect that collective decoherence conditions are
attained (also in coupled quantum dots), since all but the longest
wavelength phonon modes are unoccupied \cite{Takagahara:96}, to which
the qubits are then coupled 
collectively \cite{Zanardi:97,Lidar:PRA99Exchange}. Our results
therefore imply that 
quantum computation on DFSs in nuclear spin arrays and quantum dots should be
possible with carefully controlled exchange interactions. 

In summary, we have derived general conditions for fault tolerant quantum
computation on a DFS, and shown how to implement a universal set of gates
for the important case of collective decoherence by turning on/off
only $2$-qubit exchange Hamiltonians. In our construction the system
{\em never} 
leaves DFS during the entire execution of a gate, so that fault-tolerance is
natural and, in stark contrast to the usual situation in quantum error
correction, necessitates no extra resources during the computation. Our results are directly applicable
to any quantum computer architecture in which quantum logic is
implemented using exchange interactions, in particular to some of the recent
promising solid-state proposals for quantum computation.

Acknowledgements.--- This material is based upon work supported by the U.S.
Army Research Office under contract/grant number DAAG55-98-1-0371. We thank
Kenneth Brown for useful discussions.

\end{multicols}

\end{document}